\documentclass{Interspeech}

\usepackage{amsmath,amssymb,amsfonts}
\usepackage{algorithmic}
\usepackage{graphicx}
\usepackage{textcomp}
\usepackage{xcolor}
\def\BibTeX{{\rm B\kern-.05em{\sc i\kern-.025em b}\kern-.08em
    T\kern-.1667em\lower.7ex\hbox{E}\kern-.125emX}}
\usepackage{soul,framed} %,caption
\colorlet{shadecolor}{yellow}
\usepackage{array}
\usepackage{url}
\usepackage{cite}
\usepackage{booktabs}
\usepackage{balance} 
\usepackage{multirow}
\usepackage{multicol}
\usepackage{float}
\usepackage{bbold}
\usepackage{enumitem}
\usepackage{hyperref}
\usepackage{etoolbox,siunitx}
\robustify\bfseries
\usepackage{balance}
\usepackage{pifont}% http://ctan.org/pkg/pifont
\usepackage{scalerel}
\usepackage{caption} 

\interspeechcameraready

\title{ClearerVoice-Studio: Bridging Advanced Speech Processing Research and Practical Deployment}

\author{Shengkui}{Zhao}
\author{Zexu}{Pan}
\author{Bin}{Ma}

\affiliation{Tongyi Lab}{Alibaba Group}{Singapore}
\email{\{shengkui.zhao, zexu.pan, b.ma\}@alibaba-inc.com}

\usepackage{comment}

\begin{document}

\maketitle

\begin{abstract}
    
    This paper introduces ClearerVoice-Studio, an open-source, AI-powered speech processing toolkit designed to bridge cutting-edge research and practical application. Unlike broad platforms like SpeechBrain and ESPnet, ClearerVoice-Studio focuses on interconnected speech tasks of speech enhancement, separation, super-resolution, and multimodal target speaker extraction. A key advantage is its state-of-the-art pretrained models, including FRCRN (3M+ uses) and MossFormer (2.5M+ uses), optimized for real-world scenarios. It also offers model optimization tools, multi-format audio support, the SpeechScore evaluation toolkit, and user-friendly interfaces, catering to researchers, developers, and end-users. Its rapid adoption (2.8K GitHub stars, 200+ forks) highlights its academic and industrial impact. This paper details ClearerVoice-Studio’s \footnote{https://github.com/modelscope/ClearerVoice-Studio} capabilities, architectures, training strategies, benchmarks, community impact, and future plan.

\end{abstract}

\section{Introduction}

Speech processing is now fundamental to modern technology, powering advancements in communication, voice interfaces, and multimedia \cite{Purwins2019DeepLearning}.
%\cite{yun2017, Purwins2019DeepLearning, yan2020}.  
While crucial for these applications, accurately processing speech is challenged by the often degraded quality of real-world audio.  Noise, interfering speech, reverberation, and low resolution commonly corrupt recordings, severely impacting the performance of speech-based systems and human perception. %\cite{clement2017}
%, Chen2024RTLA, yuan2025}. 
Robust speech processing techniques are therefore essential for reliable operation in diverse and challenging environments \cite{zhang2025mambaspeech, Qiquan2020DeepMMSE}.

In recent years, robust speech processing has witnessed significant advancements, driven by cutting-edge research and innovative techniques. To democratize access to these advancements and foster collaborative innovation, several general-purpose speech open-source toolkits, such as SpeechBrain \cite{ravanelli2021} and ESPnet \cite{watanabe2018}, have emerged. These toolkits have accelerated research progress and provided essential resources for a wide range of speech-related tasks. However, they primarily focus on replicating results from research papers using public datasets, which often do not translate seamlessly to real-world deployment. In addition, there is a growing trend in multimodal speech processing, where modalities such as visual, text, or brain signals provide valuable complementary information \cite{Hang2024}. However, general toolkits like SpeechBrain and ESPnet do not yet support these multimodal approaches. Bridging these gaps requires addressing several challenges, including a unified platform design for multimodal support, the need for large-scale real-world datasets, efficient model optimization, comprehensive performance evaluations, and user-friendly interfaces, all of which are essential for enabling practical deployment and broader adoption among end users.

In addition to general-purpose toolkits, numerous open-source repositories have focused on specific speech processing tasks. For example, Asteroid \cite{Pariente2020Asteroid} specializes in audio source separation but falls short in model performance. DeepFilterNet \cite{Schroter2023DeepFilterNet} offers a lightweight speech enhancement solution but lacks broader functionality. Resemble Enhance\footnote{https://github.com/resemble-ai/resemble-enhance} targets denoising for speech quality improvement using a dual-module approach. AudioSR \cite{Liu2024audiosr} excels in audio super-resolution but struggles in noisy environments. FlowAVSE \cite{jung2024flowav} specializes in audio-visual speech enhancement, particularly for real-time applications. While these tools offer valuable contributions for research or single-task solutions, they often lack the comprehensive scope and integration needed to address the complex, interconnected challenges present in real-world scenarios. For example, a noisy recording may also contain overlapping speakers, or a low-resolution recording might be further compromised by noise. As a result, there is a strong demand for a unified toolkit capable of tackling multiple challenges within a cohesive framework. Moreover, there is a need for a unified platform that enables benchmarking of multimodal speech processing methods within the research community, facilitating the development of better models.

%There is a growing trend in multimodal speech processing, where modalities such as visual, text, or brain signals provide valuable complementary information. However, general toolkits like SpeechBrain and ESPnet do not yet support these multimodal approaches. While several open-source repositories offer implementations of research models, they often lack the robustness and usability required for practical applications. To address this gap, there is a pressing need for a unified platform that enables benchmarking of multimodal speech processing methods within the research community, facilitating the development of better models. Additionally, there is a need for state-of-the-art pretrained models trained on large-scale datasets, benefiting both developers and end-users.

To bridge these gaps, we introduce ClearerVoice-Studio, a versatile and user-friendly speech processing toolkit that leverages state-of-the-art techniques to address a wide range of real-world challenges. It seamlessly integrates solutions for speech enhancement (SE), speech separation (SS), super-resolution (SR), and multimodal target speaker extraction (AVSE) into a unified platform. By combining cutting-edge models, practical optimization and fine-tuning tools, comprehensive speech evaluation metrics, and an intuitive user-friendly interface, ClearerVoice-Studio empowers researchers, developers, and end-users to efficiently tackle complex speech processing tasks with ease and efficiency. 
%Following this introduction, Section 2 explores the toolkit’s functionalities and core model architectures. Section 3 delves into the training strategies, model optimization, and performance benchmarks. Section 4 discusses the toolkit’s community impact and adoption, as well as future development directions. Section 5 provides the conclusions.

\section{Functionalities and architectures}
% \section{ClearerVoice-Studio: functionalities and architectures}
%Detailed description of ClearerVoice-Studio's capabilities for each targeted task:
%\begin{itemize}
%\item Speech Enhancement: Available models, specific functionalities, input/output.
%\item Speech Separation: Available models, specific functionalities, input/output.
%\item Speech Super-resolution: Available models, specific functionalities, input/output.
%\item Multi-modal Target Speaker Extraction: Available models, specific functionalities, input/output, modality support.
%\end{itemize}

%In-depth explanation of the core model architectures:
%\begin{itemize}
%\item FRCRN: Architecture details, rationale for design choices.
%\item MossFormer: Architecture details, rationale for design choices, multi-modal integration.
%\item Other relevant models: Architecture details, specific advantages.
%\end{itemize}

This section details ClearerVoice-Studio's functionalities across its target tasks and provides an in-depth look at the core model architectures that power these functionalities.

\subsection{ClearerVoice-Studio functionalities}
\subsubsection{Easy access to pre-trained models}
ClearerVoice-Studio provides a suite of advanced, pre-trained models designed for a variety of SE, SS, SR, and AVSE tasks.

%\begin{itemize}
    %\item SE task: 
    \par
    \noindent
    \textbf{SE task}: ClearerVoice-Studio utilizes state-of-the-art architectures like FRCRN \cite{Zhao2022B} and MossFormer2 \cite{Zhao2024M} to deliver advanced noise reduction for speech signals.  It includes three models: FRCRN\_SE\_16K and MossFormerGAN\_SE\_16K for 16 kHz audio, and MossFormer2\_SE\_48K for 48 kHz audio. The models process noisy audio inputs, whether in raw wave format or in common formats like MP3, OGG, and AAC, producing clean, denoised outputs while supporting multiple input sampling rates. By effectively mitigating ambient noise, interference, and reverberation, these models enhance speech clarity while preserving the natural characteristics of the original audio. ClearerVoice-Studio’s speech enhancement capabilities are applicable across various scenarios, including improving the quality of meeting recordings, enhancing podcasts and audiobooks, optimizing online communication, pre-processing audio for voice cloning, conversion, or synthesis, and restoring historical recordings.
    
    %\item SS task: 
    \par
    \noindent
    \textbf{SS task}: ClearerVoice-Studio offers MossFormer2\_SS\_16K for 16 kHz robust speech separation capabilities for scenarios involving overlapping speakers.  Leveraging advanced architectures like MossFormer \cite{Zhao2023M} and MossFormer2 \cite{Zhao2024M}, the toolkit effectively decomposes mixed audio signals into distinct, speaker-specific streams.  Specifically, given a single-channel mixture containing multiple speakers, the output comprises individually separated signals for each. This functionality is crucial for applications like automatic transcription, speaker diarization, and multi-party communication systems, where accurately distinguishing and isolating individual speakers significantly improves downstream processing and overall speech intelligibility. By addressing the complexities of real-world audio environments, ClearerVoice-Studio ensures robust performance even in scenarios with high levels of overlapping speech and background noise.
    
    %\item SR task: 
    \par
    \noindent
    \textbf{SR task}: ClearerVoice-Studio also provides MossFormer2\_SR\_48K, which upscales speech audio above 16 kHz to 48 kHz for enhancing low-resolution speech by restoring lost frequency components \cite{zhao2025hifis}.  Trained on paired low- and high-resolution speech datasets, the speech super-resolution models leverage a combination architecture of MossFormer2 and HiFiGAN \cite{kong2020hifi} to accurately reconstruct missing spectral details.  Input as a low-resolution waveform yields an enhanced, high-resolution output with improved intelligibility and naturalness.  This capability is particularly valuable for restoring degraded or compressed recordings, enhancing telecommunication audio, and improving speech quality in bandwidth-limited environments.
    
    %\item AVSE task: 
    \par
    \noindent
    \textbf{AVSE task}: ClearerVoice-Studio further advances multimodal target speaker extraction by supporting a variety of conditioning modalities, including face-conditioned (AV-TFGridNet~\cite{pan2023avse}), gesture-conditioned (SEG~\cite{pan2022seg}), speech-conditioned (SpEx+~\cite{spex_plus2020}), and brain-EEG-conditioned (NeuroHeed~\cite{pan2023neuroheed}) speaker extraction. The platform provides replicated pre-trained models from respective state-of-the-art research papers, alongside a newly proposed face-conditioned speaker extraction model named AV\_MossFormer2\_TSE\_16K. This new model, trained on extensive datasets with advanced architectures, supports 16 kHz speaker isolation using synchronized face recordings. It can accurately extract a target speaker’s voice, even in challenging conditions with overlapping speech or degraded audio quality.
%\end{itemize}

These four functionalities, while available as individual modules, can be combined into composite processing pipelines.  For example, speech enhancement can precede super-resolution to address background noise before spectral detail restoration.  Similarly, super-resolution can be applied to the output of speech separation or multimodal target speaker extraction for further quality enhancement.

\subsubsection{Comprehensive training scripts}
ClearerVoice-Studio offers comprehensive training and fine-tuning scripts for its advanced speech processing suite, encompassing speech enhancement, separation, super-resolution, and multimodal target speaker extraction.  Users can easily configure hyperparameters, loss functions, and evaluation metrics for task-specific model adaptation, significantly reducing development time and resource requirements.

\subsubsection{Speech evaluation resource: SpeechScore}
ClearerVoice-Studio includes SpeechScore, a comprehensive resource for evaluating the quality and intelligibility of processed speech.  SpeechScore employs both subjective and objective evaluation methods, incorporating a range of common metrics (e.g., DNSMOS, BSSEval, PESQ, STOI, SRMR, MCD) to provide detailed insights into voice processing algorithm performance, ensuring high standards for clarity, naturalness, and overall quality.

\subsubsection{User-friendly interfaces}
ClearerVoice-Studio offers user-friendly interfaces (CLI and GUIs) for diverse users.  The CLI caters to researchers and developers needing script-based interactions, providing flexible training and inference scripts.  Web-based GUIs (HuggingFace\footnote{https://huggingface.co/spaces/alibabasglab/ClearVoice} and ModelScope\footnote{https://modelscope.cn/studios/iic/ClearerVoice-Studio} Spaces) allow non-technical users to easily upload audio and apply models with real-time feedback.  This dual approach promotes broader adoption in academic and industrial settings.

\subsection{Core model architectures}
ClearerVoice-Studio employs two core architectures: FRCRN and MossFormer2.  FRCRN\_SE\_16K uses the FRCRN architecture, while all other models are based on MossFormer2. 
%FRCRN is an attention-free network relying on complex-valued modeling and frequency recurrence for strong performance. MossFormer2 combines gated attention, convolution augmentation, and a recurrent module for enhanced modeling, and is used across multiple ClearerVoice-Studio tasks.

\subsubsection{FRCRN architecture}
FRCRN \cite{Zhao2022B} employs a convolutional-recurrent network (CRN) design, integrating a convolutional encoder-decoder (CED) with a recurrent structure.  It uses complex-valued CNNs to capture local spectral patterns in spectrograms and complex-valued RNNs (FSMNs) for temporal dependency modeling.  A key feature is its frequency recurrence mechanism, which processes frequency bands sequentially for fine-grained spectral detail capture.  Operating on time-frequency spectrograms, FRCRN predicts the complex Ideal Ratio Mask (cIRM) and has achieved strong speech enhancement results, including notable performance in the ICASSP 2022 DNS Challenge. FRCRN\_SE\_16K adopts the FRCRN architecture with parameters matching the original paper \cite{Zhao2022B}, but concatenates two FRCRN networks for enhanced mask prediction.

\subsubsection{MossFormer2 architecture}
MossFormer2 \cite{Zhao2024M} employs a hybrid design that extends the Transformer mechanism proposed in MossFormer \cite{Zhao2023M} by integrating a recurrent module. The Transformer mechanism strengths the modeling capabilities through a novel gated single-head Transformer with convolution-augmented joint self-attentions. Compared to traditional multi-head attentions,  the gated single-head Transformer is able to offer more efficient handling of both long-range interactions and local features. In addition, the joint local and global self-attention addresses long-sequence modeling by performing full self-attention on local blocks and linearized self-attention across the entire sequence. Besides the capabilities provided by the Transformer mechanism, the recurrent module is stacked to the attention block to model intricate temporal dependencies within speech signals. 

\subsubsection{Adopting MossFormer2 across tasks}
MossFormer2 serves as the primary feature mapping network architecture in ClearerVoice-Studio.  MossFormer2\_SE\_48K uses MossFormer2 to predict the phase-sensitive mask (PSM) from mel-spectrograms derived from input waveforms, applying the mask to input complex spectrograms for output.  MossFormerGAN\_SE\_16K applies MossFormer2 to 3D feature maps extracted by a time-frequency (T-F) encoder, performing simultaneous temporal and frequency attention.  A dual-decoder structure (T-F masking and spectral decoders) predicts the complex mask and spectrum, respectively, combining them via weighted sum for the final output.  MossFormer2\_SR\_48K employs a transformer-convolutional generator, combining MossFormer2 for enriched mel-spectrogram representation and HiFiGAN for upsampling and waveform reconstruction \cite{zhao2025hifis}.  MossFormer2\_SS\_16K uses a time-domain encoder-separator-decoder framework, with MossFormer2 as the masking network predicting time-domain masks and 1D convolutional encoder and decoder handling feature extraction and waveform reconstruction, respectively.
AV\_MossFormer2\_TSE\_16K is an audio-visual input extension of MossFormer2\_SS\_16K. It repeatedly fuses visual features from an additional visual encoder with speech features at the beginning of the MossFormer-Recurrent module through frame-level concatenation and projection.

\section{Training strategies and performance}
%\begin{itemize}
%\item Data preparation and augmentation techniques used for training the pre-trained models.
%\item Training procedures and optimization algorithms employed.
%\item Details on the large datasets used for training and their characteristics.
%\item Model optimization tools provided within ClearerVoice-Studio:
%\end{itemize}
\subsection{Training strategies}
\subsubsection{Dataset preparation}
ClearerVoice-Studio's pre-trained models are trained on a large and diverse dataset combining public and internal resources.  For speech enhancement, the fullband clean speech sources include the 4th DNS-Challenge speech dataset \cite{Reddy2020} and the internal TTS dataset. The 4th DNS-Challenge provides clean speech of multiple languages mainly derived from Librivox \footnote{https://librivox.org/} and VCTK corpus \cite{Yamagishi2019CSTRVC}. The audio clips were filtered based on DNS-MOS scores \cite{Reddy2020} provided by the DNS challenge organizer to control the speech quality. The VCTK corpus consists of 110 English speakers of 400 utterances each. The internal TTS dataset include 200 Mandarin speakers of 500 utterances each and 2 English speakers of 50000 utterances each. The total length of the fullband clean speech is 850 hours. 

%The fullband noise sources include the 4th DNS challenge noise dataset, the WHAM!48kHz noise dataset\footnote{http://wham.whisper.ai/}, and the internal recordings. The 4th DNS challenge noise dataset is mainly derived from the  Audioset \cite{Gemmeke2017}, Freesound\footnote{https://freesound.org/}, and  DEMAND databases \cite{thiemann_2013_1227121}. The noise clips from AudioSet were filtered using a voice activity detector to remove the clips with presence of speech. The WHAM!48kHz noise dataset primarily consists of environments from restaurants, cafes, bars, and parks. Our internal noise recordings primarily covers office and meeting room environments.  The noise data constitute 300 hours. 

Full-band noise sources include the 4th DNS Challenge noise dataset, the WHAM!48kHz noise dataset\footnote{http://wham.whisper.ai/}, and internal recordings.  The 4th DNS Challenge noise dataset is derived primarily from AudioSet \cite{Gemmeke2017}, Freesound\footnote{https://freesound.org/}, and the DEMAND dataset \cite{thiemann2013}. AudioSet noise clips were filtered using a voice activity detector to remove speech presence. The WHAM!48kHz noise dataset consists mainly of recordings from restaurants, cafes, bars, and parks, while our internal recordings cover office and meeting room environments.  The combined noise data totals 300 hours.

\begin{table}
\center
\footnotesize
\caption{Speech enhancement DNS-2020 benchmark (16 kHz). The test set contains 300 synthetic clips without reverb. SNR levels are uniformly distributed between 0 dB and 25 dB.}
\vspace{-2.5mm}
\setlength\tabcolsep{1.5pt} % default value: 6pt
\begin{tabular}{lcccc}
% \specialrule{.1em}{.05em}{.05em}
\toprule
%\multirow{2}{*}{Model} &\multirow{2}{*}{$\alpha$} &\multirow{2}{*}{$\beta$}& \multicolumn{4}{c}{Evaluation Metrics}         \\
  Model                    &{PESQ}   &{NB\_PESQ}  &{STOI}     & {P808 MOS}\\
        \midrule
        Noisy~              &1.58       &2.45       &91.52    &3.15\\
        DCCRN+~\cite{Lv2021}             &-          &3.31       &96.11    &-   \\
        MFNet~\cite{liu2023mask}              &3.43       &3.74       &97.98    &-   \\
        TridentSE~\cite{yin2022tridentse}          &3.44       &-          &97.86    &-   \\
        \midrule
        FRCRN\_SE\_16K~     &3.24       &3.66       &97.73    &4.03   \\
        MossFormerGAN\_SE\_16K  &\textbf{3.57}       &\textbf{3.88}       &\textbf{98.05}    &\textbf{4.05}\\
        %\hline
% \specialrule{.1em}{.05em}{.05em}
\bottomrule
\end{tabular}
\end{table}

\begin{table}
\center
\footnotesize
\caption{Speech enhancement VoiceBank+DEMAND benchmark (48 kHz). The test set includes 824 clips from two speakers with multiple SNR levels (-7.5 dB, -2.5 dB, 2.5 dB, and 7.5 dB)}
%\vspace{-2.5mm}
\setlength\tabcolsep{1.5pt} % default value: 6pt
\begin{tabular}{lccccc}
% \specialrule{.1em}{.05em}{.05em}
\toprule
%\multirow{2}{*}{Model} &\multirow{2}{*}{$\alpha$} &\multirow{2}{*}{$\beta$}& \multicolumn{4}{c}{Evaluation Metrics}         \\
  Model                    &{PESQ}   &{NB\_PESQ}  &{SISDR}   &{MCD}  & {P808 MOS}\\
        \midrule
        Noisy~              &1.97       &2.87       &8.39    &5.41    &3.07\\
        Resemble\_Enhance~   &2.84       &3.58       &12.42   &1.54    &\textbf{3.53}\\
        DeepFilterNet~\cite{Schroter2023DeepFilterNet}         &3.03       &3.71       &15.71   &1.77    &3.47\\
        \midrule
        MossFormer2\_SE\_48K &\textbf{3.15}  &\textbf{3.77}       &\textbf{19.36}    &\textbf{0.53}   &\textbf{3.53}\\
        %\hline
% \specialrule{.1em}{.05em}{.05em}
\bottomrule
\end{tabular}
%\vspace{-6mm}
\end{table}

Our reverberant speech was generated using 100,000 synthetic room impulse responses based on RIR\_generator\footnote{https://github.com/audiolabs/rir-generator} covering various room sizes.  Noisy-clean data was created by mixing noise with clean or reverberant speech (with 7:3 ratio) at random SNRs between 0 and 15 dB. The resulting 8000-hour training and 50-hour development sets were augmented by downsampling 10\% to 16 kHz and 5\% to 8 kHz and then upsampling to 48 kHz.  For 16 kHz wideband enhancement, the full-band data was resampled, supplemented by 262 hours of LibriTTS \cite{zen2019libritts}.  48 kHz super-resolution used the full-band clean speech, with training sets created by low-pass filtering to 16-32 kHz.  Speech separation used 107 hours of data from 110 VTCK, 202 internal TTS, and 1300 LibriTTS speakers, mixing two speakers' utterances at 0-5 dB SNR, and noise at -5 to 15 dB SNR relative to the loudest speaker.  The full-band noise and impulse responses were downsampled and reused. For AV\_MossFormer2\_TSE\_16K model, we adopt the Grid, TCDTimit, LRS2, LRS3, VoxCeleb2, MISP as speech source dataset to simulate the speech mixtures, and uses FSD50K, WHAM noise, MUSAN noise for noise sources.

\subsubsection{Model optimization}
We employed several key training strategies \cite{larochelle2009exploring} to ensure robust and efficient model development in ClearerVoice-Studio. Scalable distributed training with NCCL is used to enable multi-GPU support. Reproducibility is maintained through consistent random seeds and cuDNN settings. To facilitate multi-model training, we implemented architecture-specific model initialization and optimizer selection. Efficient data loading is achieved via distributed sampling, while gradient accumulation allows for larger effective batch sizes. Gradient clipping was adopted to enhance training stability and efficiency. We utilized a learning rate schedule with fine-tuning and halving and enforced regular check-pointing and logging for monitoring and resuming training.

%We adopted several key training strategies to ensure the robust and efficient model development  in ClearerVoice-Studio as follows. We applied scalable distributed training with NCCL for multi-GPU support\cite{Kempner_Handbook_2025}. We ensure the reproducibility via consistent random seeds and cuDNN settings \cite{zhuang2021randomness}. To support multi-model training, we developed architecture-specific model initialization and optimizer selection \cite{JMLRv25230166}. The efficient data loading is ensured with the distributed sampling method. In addition, we provided gradient accumulation scheme for effective larger batch sizes and gradient clipping for training efficiency and stability. We used a learning rate schedule with fine-tuning and halving. We further adopted gradient clipping for stability and regular checkpointing/logging for monitoring and resuming training \cite{larochelle2009exploring}. 

The training script uses different loss functions tailored to each model. FRCRN\_SE\_16K combines SiSNR and the complex-valued masking MSE loss for spectral fidelity and temporal consistency \cite{Zhao2022B}. Mossformer2\_SE\_48K uses the masking MSE loss \cite{Erdogan2015}, considering both magnitude and phase. MossformerGan\_SE\_16K employs an adversarial approach, combining generator loss (MSE on discriminator output), magnitude loss (MSE on spectrograms), real/imaginary loss (MSE on STFT coefficients), time domain loss, and discriminator loss incorporating PESQ for perceptual alignment \cite{Cao_2022}. These diverse loss functions guide model training toward optimal performance across multiple audio quality dimensions. MossFormer2\_SS\_16K uses Permutation Invariant Training (PIT) with SI-SNR loss \cite{yu2017pit}. The optimization for MossFormer2\_SR\_48K combines GAN loss corresponding to the three discriminator types of MSD, MPD, and MBD, multi-scale mel-spectrogram loss, and feature matching loss \cite{zhao2025hifis}. 

\begin{table}
\center
\footnotesize
\caption{Speech separation benchmarks (8 kHz \& 16 kHz), each containing 3000 utterances: LRS2-2Mix (noisy and reverberant, SNR -5 dB to 5 dB), WSJ0-2Mix (clean), Libri2Mix (noisy, LUFS -25 dB to -33 dB), and WHAM! (noisy, SNR -6 dB to 3 dB).}
\vspace{-2.5mm}
\setlength\tabcolsep{1.5pt} % default value: 6pt
\begin{tabular}{lcccc}
% \specialrule{.1em}{.05em}{.05em}
\toprule
%\multirow{2}{*}{Model} &\multirow{2}{*}{$\alpha$} &\multirow{2}{*}{$\beta$}& \multicolumn{4}{c}{Evaluation Metrics}         \\
  Model                    &{LRS2-2Mix}   &{WSJ0-2Mix}  &{Lirbi2Mix}   &{WHAM!} \\
        \midrule
        DualPathRNN~\cite{Luo2020Z}      &12.7       &18.8       &16.1    &13.7   \\
        SepFormer~\cite{Subakan2021M}        &13.5       &20.4       &17.0    &14.4   \\
        TDANet~\cite{tdanet2023iclr}           &14.2       &18.5       &17.4   &15.2   \\
        TF-GridNet~\cite{wang2023tf}        & -         &\textbf{22.8}       &19.8   &16.9   \\
        SPMamba~\cite{li2024spmambastatespacemodelneed}           & -         &22.5       &\textbf{19.9}   &\textbf{17.4}   \\
        \midrule
        MossFormer2\_SS\_16K &\textbf{15.5}    &22.0       &16.7    &\textbf{17.4}  \\
        %\hline
% \specialrule{.1em}{.05em}{.05em}
\bottomrule
\vspace{-2.5mm}
\end{tabular}

\end{table}
\begin{table}
\center
\footnotesize
\caption{Speech super-resolution performance evaluation. }
\vspace{-2.5mm}
\setlength\tabcolsep{1.5pt} % default value: 6pt
\begin{tabular}{lccccc}
% \specialrule{.1em}{.05em}{.05em}
\toprule
%\multirow{2}{*}{Model} &\multirow{2}{*}{$\alpha$} &\multirow{2}{*}{$\beta$}& \multicolumn{4}{c}{Evaluation Metrics}         \\
  Model            &{PESQ}   &{16 kHz}  &{24 kHz}   &{32 kHz} &{48 kHz} \\
        \midrule
        Origin~      &1.97       &2.80    &2.60    &2.29     &1.46   \\
        Enhanced     &\textbf{3.15}    &\textbf{1.93}       &\textbf{1.52}    &\textbf{1.50}      &\textbf{1.42}  \\
        %\hline
% \specialrule{.1em}{.05em}{.05em}
\bottomrule
\end{tabular}
\end{table}

\subsection{Model performance}
We provide the evaluation performance for each model below.

\par
\noindent
\textbf{SE:}
Our 16 kHz speech enhancement models are compared against DCCRN+ \cite{Lv2021}, MFNet \cite{liu2023mask}, and TridentSE \cite{yin2022tridentse} on the DNS Challenge 2020 benchmark, while our 48 kHz model is benchmarked against Resemble\_Enhance and DeepFilterNet on VoiceBank+DEMAND.  Results are shown in Tables 1 and 2, demonstrating competitive performance despite training on general-purpose datasets. 

\begin{table}
    \centering
    \sisetup{
    detect-weight, % Make siunitx detect align bold cells correctly
    mode=text, % Make siuntix print tables in text mode (causes width of bold characters to be the same as non-bold)
    tight-spacing=true,
    round-mode=places,
    round-precision=1,
    table-format=2.1
    }
    \caption{Face-conditioned speaker extraction performance on VoxCeleb2 benchmark, results are reported in dB on 2 speakers mixtures (2-mix) and 3 speakers mixtures (3-mix).}
    \vspace{-1mm}
    \addtolength{\tabcolsep}{-2pt}
    \resizebox{\linewidth}{!}{
    \begin{tabular}{c S|SS|SS}
       \toprule
        \multirow{2}*{Model}           &{\multirow{2}*{Params}} &\multicolumn{2}{c|}{2-mix} &\multicolumn{2}{c}{3-mix} \\
        &&{SI-SNRi}  &{SNRi} &{SI-SNRi}  &{SNRi} \\
        \midrule
        AV-ConvTasNet~\cite{wu2019time}   &20.2       &10.6       &10.9   &9.8 &10.2  \\
        MuSE~\cite{pan2020muse}            &25.0       &11.7       &12.0   &11.6    &12.2  \\
        reentry~\cite{pan2021reentry}         &18.8       &12.6       &12.9  &12.6 &13.1    \\
        AV-DPRNN~\cite{usev21}        &\textbf{15.3}       &11.5       &11.8  &10.5 &11.0  \\
        AV-TFGridNet~\cite{pan2023avse}    &20.8       &13.7       &14.1  &14.2    &14.6  \\
        \midrule
        AV-MossFormer2 &68.5 &\textbf{14.6}   &\textbf{14.9}&\textbf{15.5}   &\textbf{16.0}\\
        \bottomrule
    \end{tabular}
    }
    \addtolength{\tabcolsep}{2pt}
    \vspace*{-3mm}
    \label{tab:voxceleb2_2spk}
\end{table}

\par
\noindent
\textbf{SS:}
Speech separation performance was evaluated on LRS2-2Mix (16 kHz), WSJ0-2Mix (8 kHz), Libri2Mix (8 kHz), and WHAM! (8 kHz) benchmarks, comparing against DualPathRNN \cite{Luo2020Z}, SepFormer \cite{Subakan2021M}, TDANet \cite{tdanet2023iclr}, TF-GridNet \cite{wang2023tf}, and SPMamba \cite{li2024spmambastatespacemodelneed} models (results from published works). MossFormer2\_SS\_16K, evaluated on resampled 16 kHz audio from our unified model (without dataset-specific retraining), achieved competitive results on LRS2-2Mix, WSJ0-2Mix, and WHAM! as shown in Table 3. However, its lower performance on Libri2Mix may be due to the dataset's inherently lower speech quality.

\par
\noindent
\textbf{SR:}
We demonstrated the effectiveness of our speech super-resolution model, MossFormer2\_SR\_48K, using the VoiceBank+DEMAND 48 kHz test set. For super-resolution evaluation, the test set was downsampled to 16 kHz, 24 kHz, and 32 kHz. Recognizing that speech quality is impacted by both lower sampling rates and background noise, we also incorporated our speech enhancement model, MossFormer2\_SE\_48K, to reduce noise prior to super-resolution processing. Results are presented in Table 4. While MossFormer2\_SE\_48K significantly improved the 16 kHz PESQ score, its impact on LSD was marginal (see the last column).  Consequently, the LSD improvements observed at 16 kHz, 24 kHz, and 32 kHz are primarily due to MossFormer2\_SR\_48K.

\par
\noindent
\textbf{AVSE:}
%While AV-TFGridNet represents the current state-of-the-art in face-conditioned speaker extraction, we propose a new model, AV-MossFormer2, leveraging the superior performance of the MossFormer2 architecture. To standardize AVSE model evaluation, we introduce performance benchmarks on the VoxCeleb2 mixture dataset, as detailed in Table 5. Additionally, we have established a benchmark on the LRS2 mixture dataset, which is available in the repository. We further extend our benchmarking efforts to include gesture-conditioned speaker extraction on the YGD mixture dataset and brain-EEG-conditioned speaker extraction on the KUL mixture dataset, both of which are available in the repository. Additionally, pre-trained checkpoints for these models are also provided, enabling researchers and developers to easily replicate and build upon our work.
To standardize audio-visual speaker extraction (AVSE) evaluation, we introduced the VoxCeleb2 mixture dataset, and established benchmarks shown in Table 5.  AV-MossFormer2 outperforms the state-of-the-art AV-TFGridNet on both 2-speaker and 3-speaker mixtures.  Benchmarks for gesture-conditioned (YGD) and brain-EEG-conditioned (KUL) speaker extraction are also provided in the repository.

\section{Discussion and future plan}
We presented ClearerVoice-Studio serving as a valuable bridge between advanced speech processing research and practical applications. Beyond the presented evaluations, ClearerVoice-Studio is available for live demos on HuggingFace and ModelScope, enabling users to experiment with real-world recordings.  We have also generated a 48 kHz version of the LJSpeech-1.1 TTS corpus based on ClearerVoice-Studio, now publicly available\footnote{https://huggingface.co/datasets/alibabasglab/LJSpeech-1.1-48kHz}.  
ClearerVoice-Studio has rapidly gained traction within the research and development community, evidenced by over 2,800 GitHub stars and 200+ forks, signifying widespread interest and active adaptation.  
%This popularity is driven by several key factors: ease of use (straightforward APIs and comprehensive documentation), state-of-the-art performance (leveraging models like FRCRN and MossFormer with over 3M and 2.5M uses respectively on ModelScope), flexibility and extensibility, and active maintenance and support.  Consequently, ClearerVoice-Studio is being applied in both academic research and industrial settings. 

ClearerVoice-Studio's future development will focus on integrating cutting-edge models and algorithms (e.g., diffusion models), expanding support for additional tasks and modalities (e.g., multi-channel audio and video), and enhancing training tools through streamlined pipelines, improved visualization, and automated hyperparameter tuning. The development will also prioritize real-time processing and edge deployment, while continuing to foster community engagement. These initiatives will ensure ClearerVoice-Studio remains a dynamic and influential resource within the speech processing community.

% \scriptsize
\tiny
\bibliographystyle{IEEEtran}
\bibliography{mybib}

\begin{thebibliography}{10}
\providecommand{\url}[1]{#1}
\csname url@samestyle\endcsname
\providecommand{\newblock}{\relax}
\providecommand{\bibinfo}[2]{#2}
\providecommand{\BIBentrySTDinterwordspacing}{\spaceskip=0pt\relax}
\providecommand{\BIBentryALTinterwordstretchfactor}{4}
\providecommand{\BIBentryALTinterwordspacing}{\spaceskip=\fontdimen2\font plus
\BIBentryALTinterwordstretchfactor\fontdimen3\font minus
  \fontdimen4\font\relax}
\providecommand{\BIBforeignlanguage}[2]{{%
\expandafter\ifx\csname l@#1\endcsname\relax
\typeout{** WARNING: IEEEtran.bst: No hyphenation pattern has been}%
\typeout{** loaded for the language `#1'. Using the pattern for}%
\typeout{** the default language instead.}%
\else
\language=\csname l@#1\endcsname
\fi
#2}}
\providecommand{\BIBdecl}{\relax}
\BIBdecl

\bibitem{Purwins2019DeepLearning}
H.~Purwins, B.~Li, T.~Virtanen, J.~Schl{\"u}ter, S.-Y. Chang, and T.~Sainath,
  ``Deep learning for audio signal processing,'' \emph{IEEE J. Sel. Top. Signal
  Process.}, vol.~13, no.~2, pp. 206--219, 2019.

\bibitem{zhang2025mambaspeech}
X.~Zhang, Q.~Zhang, H.~Liu, T.~Xiao, X.~Qian, B.~Ahmed, E.~Ambikairajah, H.~Li,
  and J.~Epps, ``Mamba in {S}peech: Towards an alternative to self-attention,''
  \emph{preprint arXiv:2405.12609}, 2025.

\bibitem{Qiquan2020DeepMMSE}
Q.~Zhang, A.~Nicolson, M.~Wang, K.~K. Paliwal, and C.~Wang, ``Deep{MMSE}: A
  deep learning approach to mmse-based noise power spectral density
  estimation,'' \emph{IEEE/ACM Trans. Audio Speech Lang. Process.}, vol.~28,
  pp. 1404--1415, 2020.

\bibitem{ravanelli2021}
\BIBentryALTinterwordspacing
M.~Ravanelli, T.~Parcollet, P.~Plantinga, A.~Rouhe, S.~Cornell, L.~Lugosch,
  C.~Subakan, N.~Dawalatabad, A.~Heba, J.~Zhong, J.-C. Chou, S.-L. Yeh, S.-W.
  Fu, C.-F. Liao, E.~Rastorgueva, F.~Grondin, W.~Aris, H.~Na, Y.~Gao, R.~D.
  Mori, and Y.~Bengio, ``Speech{B}rain: A general-purpose speech toolkit,''
  2021. [Online]. Available: \url{https://arxiv.org/abs/2106.04624}
\BIBentrySTDinterwordspacing

\bibitem{watanabe2018}
S.~Watanabe, T.~Hori, S.~Karita, T.~Hayashi, J.~Nishitoba, Y.~Unno, N.~E.~Y.
  Soplin, J.~Heymann, M.~Wiesner, N.~Chen, A.~Renduchintala, and T.~Ochiai,
  ``{ESP}net: End-to-end speech processing toolkit,'' in \emph{Proc.
  Interspeech}, 2018.

\bibitem{Hang2024}
H.~Chen, S.~Wu, C.~Wang, J.~Du, C.-H. Lee, S.~M. Siniscalchi, S.~Watanabe,
  J.~Chen, O.~Scharenborg, Z.-Q. Wang, B.-C. Yin, and J.~Pan, ``Summary on the
  multimodal information-based speech processing 2023 challenge,'' in
  \emph{Proc. ICASSPW}, 2024, pp. 123--124.

\bibitem{Pariente2020Asteroid}
M.~Pariente, S.~Cornell, J.~Cosentino, S.~Sivasankaran, E.~Tzinis,
  J.~Heitkaemper, M.~Olvera, F.-R. Stöter, M.~Hu, J.~M. Martín-Doñas,
  D.~Ditter, A.~Frank, A.~Deleforge, and E.~Vincent, ``Asteroid: the
  {PyTorch}-based audio source separation toolkit for researchers,'' in
  \emph{Proc. Interspeech}, 2020.

\bibitem{Schroter2023DeepFilterNet}
H.~Schröter, A.~N. Escalante, , T.~Rosenkranz, and A.~Maier,
  ``Deep{F}ilter{N}et: Perceptually motivated real-time speech enhancement,''
  in \emph{Proc. Interspeech}, 2023.

\bibitem{Liu2024audiosr}
H.~Liu, K.~Chen, Q.~Tian, W.~Wang, and M.~D. Plumbley, ``Audio{SR}: Versatile
  audio super-resolution at scale,'' in \emph{Proc. ICASSP}, 2024, pp.
  1076--1080.

\bibitem{jung2024flowav}
C.~Jung, S.~Lee, J.-H. Kim, and J.~S. Chung, ``Flow{AVSE}: Efficient
  audio-visual speech enhancement with conditional flow matching,'' in
  \emph{Proc. Interspeech}, 2024.

\bibitem{Zhao2022B}
S.~Zhao, B.~Ma, K.~N. Watcharasupat, and W.-S. Gan, ``{FRCRN}: {B}oosting
  feature representation using frequency recurrence for monaural speech
  enhancement,'' in \emph{Proc. ICASSP}, 2022.

\bibitem{Zhao2024M}
S.~Zhao, Y.~Ma, C.~Ni, C.~Zhang, H.~Wang, T.~H. Nguyen, K.~Zhou, J.~Yip, D.~Ng,
  and B.~Ma, ``Moss{F}ormer2: Combining transformer and rnn-free recurrent
  network for enhanced time-domain monaural speech separation,'' in \emph{Proc.
  ICASSP}, 2024.

\bibitem{Zhao2023M}
S.~Zhao and B.~Ma, ``Moss{F}ormer: Pushing the performance limit of monaural
  speech separation using gated single-head transformer with
  convolution-augmented joint self-attentions,'' in \emph{Proc. ICASSP}, 2023.

\bibitem{zhao2025hifis}
S.~Zhao, K.~Zhou, Z.~Pan, Y.~Ma, C.~Zhang, and B.~Ma, ``Hifi-{SR}: A unified
  generative transformer-convolutional adversarial network for high-fidelity
  speech super-resolution,'' in \emph{Proc. ICASSP}, 2025.

\bibitem{kong2020hifi}
J.~Kong, J.~Kim, and J.~Bae, ``Hi{F}i-{GAN}: Generative adversarial networks
  for efficient and high fidelity speech synthesis,'' in \emph{Proc. NeurIPS},
  2020.

\bibitem{pan2023avse}
Z.~Pan, G.~Wichern, Y.~Masuyama, F.~G. Germain, S.~Khurana, C.~Hori, and
  J.~Le~Roux, ``Scenario-aware audio-visual {TF-Gridnet} for target speech
  extraction,'' in \emph{Proc. ASRU}, 2023.

\bibitem{pan2022seg}
Z.~Pan, X.~Qian, and H.~Li, ``Speaker extraction with co-speech gestures cue,''
  \emph{IEEE Signal Process. Lett.}, vol.~29, pp. 1467--1471, 2022.

\bibitem{spex_plus2020}
M.~Ge, C.~Xu, L.~Wang, E.~S. Chng, J.~Dang, and H.~Li, ``{SpEx}+: A complete
  time domain speaker extraction network,'' in \emph{Proc. Interspeech}, 2020,
  pp. 1406--1410.

\bibitem{pan2023neuroheed}
Z.~Pan, M.~Borsdorf, S.~Cai, T.~Schultz, and H.~Li, ``Neuro{H}eed:
  Neuro-steered speaker extraction using {EEG} signals,''
  \emph{arXiv:2307.14303}, 2023.

\bibitem{Reddy2020}
C.~K. Reddy, V.~Gopal, R.~Cutler, E.~Beyrami, R.~Cheng, H.~Dubey,
  S.~Matusevych, R.~Aichner, A.~Aazami, and S.~B. {\it et al}., ``The
  interspeech 2020 deep noise suppression challenge: Datasets, subjective
  testing framework, and challenge results,'' in \emph{Proc. Interspeech},
  2020.

\bibitem{Yamagishi2019CSTRVC}
\BIBentryALTinterwordspacing
J.~Yamagishi, C.~Veaux, and K.~MacDonald, ``{CSTR VCTK C}orpus: English
  multi-speaker corpus for cstr voice cloning toolkit (version 0.92),'' 2019.
  [Online]. Available: \url{https://api.semanticscholar.org/CorpusID:213060286}
\BIBentrySTDinterwordspacing

\bibitem{Gemmeke2017}
J.~F. Gemmeke, D.~P.~W. Ellis, D.~Freedman, A.~Jansen, W.~Lawrence, R.~C.
  Moore, M.~Plakal, and M.~Ritter, ``Audio set: An ontology and human-labeled
  dataset for audio events,'' in \emph{Proc. ICASSP}, 2017, pp. 776--780.

\bibitem{thiemann2013}
\BIBentryALTinterwordspacing
J.~Thiemann, N.~Ito, and E.~Vincent, ``{DEMAND}: a collection of multi-channel
  recordings of acoustic noise in diverse environments,'' Jun. 2013. [Online].
  Available: \url{https://doi.org/10.5281/zenodo.1227121}
\BIBentrySTDinterwordspacing

\bibitem{Lv2021}
S.~Lv, Y.~Hu, S.~Zhang, and L.~Xie, ``{DCCRN}+: Channel-wise subband dccrn with
  snr estimation for speech enhancement,'' in \emph{Proc. Interspeech}, 2021.

\bibitem{liu2023mask}
L.~Liu, H.~Guan, J.~Ma, W.~Dai, G.~Wang, and S.~Ding, ``A mask free neural
  network for monaural speech enhancement,'' in \emph{Proc. Interspeech}, 2023.

\bibitem{yin2022tridentse}
D.~Yin, Z.~Zhao, C.~Tang, Z.~Xiong, and C.~Luo, ``Trident{SE}: Guiding speech
  enhancement with 32 global tokens,'' in \emph{Proc. Interspeech}, 2023.

\bibitem{zen2019libritts}
H.~Zen, V.~Dang, R.~Clark, Y.~Zhang, R.~J. Weiss, Y.~Jia, Z.~Chen, and Y.~Wu,
  ``Libri{TTS}: A corpus derived from librispeech for text-to-speech,'' in
  \emph{Proc. Interspeech}, 2019.

\bibitem{larochelle2009exploring}
H.~Larochelle, Y.~Bengio, J.~Louradour, and P.~Lamblin, ``Exploring strategies
  for training deep neural networks,'' \emph{Journal of Machine Learning
  Research}, vol.~10, no.~1, pp. 1--40, 2009.

\bibitem{Erdogan2015}
H.~Erdogan, J.~R. Hershey, S.~Watanabe, and J.~Le~Roux, ``Phase-sensitive and
  recognition-boosted speech separation using deep recurrent neural networks,''
  in \emph{Proc. ICASSP}, 2015, pp. 708--712.

\bibitem{Cao_2022}
R.~Cao, S.~Abdulatif, and B.~Yang, ``{CMGAN}: Conformer-based metric gan for
  speech enhancement,'' in \emph{Proc. Interspeech}, 2022.

\bibitem{yu2017pit}
D.~Yu, M.~Kolbæk, Z.-H. Tan, and J.~Jensen, ``Permutation invariant training
  of deep models for speaker-independent multi-talker speech separation,'' in
  \emph{Proc. ICASSP}, 2017.

\bibitem{Luo2020Z}
Y.~Luo, Z.~Chen, and T.~Yoshioka, ``{D}ual-{P}ath {RNN}: {E}fficient long
  sequence modeling for time-domain single-channel speech separation,'' in
  \emph{Proc. ICASSP}, 2020.

\bibitem{Subakan2021M}
C.~Subakan, M.~Ravanelli, S.~Cornell, M.~Bronzi, and J.~Zhong, ``{A}ttention is
  all you need in speech separation,'' in \emph{Proc. ICASSP}, 2021.

\bibitem{tdanet2023iclr}
K.~Li, R.~Yang, and X.~Hu, ``An efficient encoder-decoder architecture with
  top-down attention for speech separation,'' in \emph{Proc. ICLR}, 2023.

\bibitem{wang2023tf}
Z.-Q. Wang, S.~Cornell, S.~Choi, Y.~Lee, B.-Y. Kim, and S.~Watanabe,
  ``{TF-G}rid{N}et: Making time-frequency domain models great again for
  monaural speaker separation,'' in \emph{Proc. ICASSP}, 2023.

\bibitem{li2024spmambastatespacemodelneed}
K.~Li, G.~Chen, R.~Yang, and X.~Hu, ``{SPM}amba: State-space model is all you
  need in speech separation,'' \emph{arXiv:2404.02063}, 2024.

\bibitem{wu2019time}
J.~{Wu}, Y.~{Xu}, S.~{Zhang}, L.~{Chen}, M.~{Yu}, L.~{Xie}, and D.~{Yu}, ``Time
  domain audio visual speech separation,'' in \emph{Proc. ASRU}, 2019.

\bibitem{pan2020muse}
Z.~Pan, R.~Tao, C.~Xu, and H.~Li, ``{MuSE}: Multi-modal target speaker
  extraction with visual cues,'' in \emph{Proc. ICASSP}, 2021.

\bibitem{pan2021reentry}
------, ``Selective listening by synchronizing speech with lips,''
  \emph{IEEE/ACM Trans. Audio, Speech, Lang. Process.}, vol.~30, pp.
  1650--1664, 2022.

\bibitem{usev21}
Z.~Pan, M.~Ge, and H.~Li, ``{USEV}: Universal speaker extraction with visual
  cue,'' \emph{IEEE/ACM Trans. Audio, Speech, Lang. Process.}, vol.~30, pp.
  3032--3045, 2022.

\end{thebibliography}

\end{document}